\documentstyle[preprint,aps]{revtex}
\begin{document}
\draft
\newfont{\form}{cmss10}
\newcommand{\unity}{1\kern-.65mm \mbox{\form l}}
\newcommand{\k}{\mbox{\form l}\kern-.6mm \mbox{\form K}}
\newcommand{\D}{D \raise0.5mm\hbox{\kern-2.0mm /}}
\newcommand{\A}{A \raise0.5mm\hbox{\kern-1.8mm /}}
\def\pmb#1{\leavevmode\setbox0=\hbox{$#1$}\kern-.025em\copy0\kern-\wd0
\kern-.05em\copy0\kern-\wd0\kern-.025em\raise.0433em\box0}
\title{Yang-Mills theories on the space-time $S_1\times R$ cylinder:
equal-time quantization in light-cone gauge and Wilson loops}
\author{A. Bassetto}
\address{Dipartimento di Fisica ``G.Galilei", Via Marzolo 8 -- 35131
Padova, Italy \\ INFN, Sezione di Padova, Italy}
\author{L. Griguolo}
\address{Dipartimento di Fisica ``G.Galilei", Via Marzolo 8 -- 35131
Padova, Italy \\ Center for Theoretical Physics, Laboratory of
Nuclear Science, MIT, Cambridge, MA 02139, USA}
\author{G. Nardelli}
\address{Dipartimento di Fisica , Universit\`a di Trento and
\\ INFN, Gruppo Collegato di Trento, 38050 Povo (Trento), Italy}
\maketitle
\begin{abstract}
Pure Yang-Mills theories on the $S_1\times R$ cylinder are
quantized in light-cone gauge $A_-=0$ by means of
${\bf equal-time}$ commutation relations.
Positive and negative frequency components are excluded
from the ``physical" Hilbert space by imposing Gauss' law
in a weak sense. Zero modes, related
to the winding on the cylinder, provide non trivial
topological variables of the theory. A Wilson loop with light-like
sides is studied:
in the abelian case it can be exactly computed obtaining the
expected area result, whereas difficulties are pointed
out in non abelian cases.
\end{abstract}
\pacs{11.10Kk,11.15.Bt \qquad DFPD 95/TH/07}

\narrowtext
\section{Introduction}
\label{prima}
Yang-Mills theories on the two-dimensional space-time cylinder
$S_1\times R$ have been extensively studied in the recent
literature \cite{Ra88,He88,He93,He94}. The common wisdom says that, without
fermions, no dynamical
degrees of freedom are available in two dimensions, but only
topological excitations which, to be studied, need (at least
a partial) compactification of the base manifold. As we have
in mind the possibility of studying Wilson loops with light-like
sides, we shall strictly comply with a Minkowski formulation and
put our system in a
spatial box with periodic boundary conditions. The system
will be allowed to evolve in time according
to its peculiar dynamics. As for the
fiber, we shall confine ourselves to $SU(N)$, although the
treatment might be further generalized. In passing,
also the simpler case of QED will be discussed.
We shall consider
pure Yang-Mills theories, deferring to a future investigation
the introduction of dynamical
fermions, which would entail important consequences on the Hilbert
space structure of the theory.

Quantization has been performed in the literature
using Coulomb or Hamilton
gauges on the group algebra \cite{He88}; the toroidal structure
of the system has been fully clarified, together with the
properties of Gribov's copies \cite{He93}. In particular the
energy levels and the Hilbert space of states have been
obtained.

We should also mention the possibility of quantizing the system
on the group manifold itself (and not on its algebra) \cite{Ra88}.
In this case one can avoid gauge fixing, but the spectrum
one gets is different from the one obtained in the previous
approach by the Casimir ``energy" due to the group curvature \cite{He94}.

In the following we shall restrict ourselves to the group algebra
approach.
We are interested in learning what are the differences occurring
when the system is quantized in non-covariant light-cone gauge.
As is well known, this gauge has advantages and difficulties on its
own. First of all it is not allowed to compactify
the base manifold in the gauge vector direction; in the case we are
studying the condition $A_1=0$ would lead to inconsistencies. No
contradiction will instead ensue from the choice $A_-=0$; we shall
indeed refrain from using light-front quantization: to compactify
the range of $x^-$ would not be allowed without drastically changing
the gauge condition ($e.g.$ by choosing $\partial_-A_-=0$). We shall
quantize our system according to equal time commutation relations.

There are several reasons supporting this choice: from a physical
view point a system naturally evolves in ${\bf time}$. To force
evolution in $x^+$ entails the appearance of unnecessary zero modes
which in turn generate ``artificial" infrared singularities.
On the other hand no simplification would ensue, as we want to
put the system in a spatial box $-L\le x \le L$, with periodic
boundary conditions.

At variance with the Coulomb gauge choice,
the Faddeev-Popov determinant is irrelevant in this case.

Our interest in light-cone gauge on a cylinder is prompted by some
results we have found in the continuum in strictly $1+1$
dimensions \cite{Basgri94}. To summarize them,
we recall that different finite expressions are obtained for a rectangular
Wilson loop with light-like sides, of ``lengths" $\lambda$ and $\tau$
respectively, according to different choices
for the vector propagator.
The first choice has the ``contact" form of a real
delta-like interaction concentrated at the value $x+t=0$ (see
$Sect$.III). It has
been suggested by 't Hooft \cite{Ho74} and used in performing
the calculation of the quark propagator and of the mesonic wave
functions in the large-N limit. However, as was already noticed at that time
\cite{Wu77}, this propagator exhibits extra singularities under
Wick's rotation.

Canonical equal time quantization in the continuum suggests instead
a different expression for the propagator in which massless
excitations contribute in a causal way, although with negative
probability (ghosts)\cite{Ba85,Basgri94}.
They give rise indeed to a complex ``potential"
with an absorptive part, which possesses smooth properties under Wick's
rotation.
This form can also be obtained as the two-dimensional limit of the
$d$-dimensional Mandelstam-Leibbrandt (M-L)
propagator, that turns out to be essential
in order to prove the renormalizability of
the theory in $d=3+1$ \cite{Bass91}.
Precisely this propagator is also required
to recover gauge invariance of the
Wilson loop in $(3-\epsilon)+1$ dimensions\cite{Bass93}.

In turn both Wilson loop results are different from the gauge
invariant expression one obtains when the theory is considered
in $(1+\epsilon)+1$ dimensions and the limit $\epsilon \to 0$
is eventually performed \cite{Basgri94}.
In our opinion this phenomenon reflects
a basic ``discontinuity" of the theory in the limit $\epsilon \to 0$,
rather than rising doubts on light-cone gauge.

As the calculation in strictly 1+1 dimensions would exhibit in the continuum
infrared divergences when choosing
Feynman or Coulomb gauges, at least in intermediate steps, and thereby
would prevent a direct comparison with light-cone gauge results, as a first
step we feel interesting a careful understanding of the theory on a cylinder
when quantized in light-cone gauge imposing equal $time$ canonical
commutation relations.

This is done in $Sect$.II,
paying particular attention both to zero modes and to positive
and negative frequency excitations. Canonical quantization,
Gauss's law , Hilbert space of states and eigenvalues of the
hamiltonian are carefully discussed for the algebras $su(2)$
and $su(3)$. In particular the spectrum of topological excitations
coincides with known results obtained in Coulomb gauge. In $Sect$.III
a particular Wilson loop with light-like sides is studied:
in the abelian case it can be exactly computed obtaining the
expected area result, whereas several difficulties are pointed
out preventing a full solution in the non abelian cases and rising
serious criticism concerning perturbative approaches.

Final conclusions are drawn and possible future developments are
suggested in
$Sect$.IV.

\vskip 0.3truecm
\section{Equal-time quantization in light-cone gauge}
\label{seconda}
\vskip 0.3truecm

We start from the lagrangian density
\begin{equation}
\label{uno}
{\cal L} = -{1 \over 2}Tr\big( F_{\mu \nu} F^{\mu \nu}\big)-2
Tr\big(\lambda nA\big).
\end{equation}

$F_{\mu \nu}$ is the usual field tensor, $A_\mu$ the vector potential,
$\lambda$ is a Lagrange multiplier and the gauge vector $n_{\mu}$ is
lightlike. The usual normalization

\begin{equation}
\label{mille}
Tr(T_{a}T_{b})={1\over 2}\delta_{ab} \nonumber
\end{equation}

is understood.
We introduce $\pm$ components and, to be specific, we choose
$n_+=1$. In the following a ``conjugate" lightlike vector $n^*_-=1$
will also be considered.

{}From the lagrangian (\ref{uno}) one can immediately derive the
Euler equations

\begin{eqnarray}
\label{due}
D^{\mu}F_{\mu\nu}&=& \lambda n_{\nu}, \nonumber\\
nA&=&0,
\end{eqnarray}

$D^{\mu}$ being the covariant derivative acting on the adjoint
representation. Eqs.(\ref{due}) can be rewritten as

\begin{equation}
\label{tre}
\partial_-^2 A_+=0,
\end{equation}

\begin{equation}
\label{quattro}
\partial_-\partial_+A_+ -ig [A_+,\partial_-A_+]=\lambda,
\end{equation}

\begin{equation}
\label{cinque}
A_-=0.
\end{equation}

The Lagrange multiplier is equal to Gauss' operator and satisfies
the equation

\begin{equation}
\label{sei}
\partial_-\lambda=0
\end{equation}

as a consequence of eqs.(\ref{tre})-(\ref{cinque}).

We consider our theory on a space-time cylinder, the space variable $x$
taking values between $-L\le x \le L$ and all quantities satisfying
periodic boundary conditions; we stress that the time variable $t$ is not
compactified. The possibility of considering the theory on a
cylinder is not trivial when imposing axial gauges: for
instance the gauge fixing $A_1=0$ would
not be allowed, as is
well known.

It is natural to expand the potential in a Fourier series

\begin{equation}
\label{sette}
A_+(t,x)\equiv A(t,x)={1\over \sqrt {2L}}
\sum_{k=-\infty}^{\infty}A_k(t)exp(i{\pi kx\over L}),
\end{equation}

\begin{equation}
\label{otto}
A_k(t)={1\over \sqrt {2L}} \int_{-L}^L A(t,x)exp(-i{\pi kx
\over L})dx.
\end{equation}

Eq.(\ref{tre}) has the solution

\begin{equation}
\label{nove}
A_k(t)=(a_kt+b_k)exp(i{\pi kt\over L})
\end{equation}

and therefrom

\begin{equation}
\label{dieci}
A(t,x)={1\over \sqrt {2L}}\sum_{k=-\infty}^{\infty}(a_kt+b_k)exp(i{\pi
k(t+x)\over L})={1\over \sqrt {2L}}\big(b_0+a_0t\big)+\hat{A}(t,x).
\end{equation}

Similarly eq.(\ref{sei}) has the solution

\begin{equation}
\label{undici}
\lambda(t,x)={1\over \sqrt {2L}}\sum_{k=-\infty}^{\infty}\lambda_k exp(i{\pi
k(t+x)\over L}).
\end{equation}

In this treatment $A$, $F\equiv F_{-+}$ and $\lambda$ look
essentially as free
fields, all dynamics being transferred to the Gauss' law
(eq.(\ref{quattro}))

\begin{equation}
\label{dodici}
\lambda_k={i\pi k\over L}a_k - {ig \over {2\sqrt{L}}}
\sum_{p,q=-\infty}^{\infty}[b_p,a_q]\delta_{k,p+q}.
\end{equation}

One can also easily derive the hamiltonian corresponding to
(\ref{uno})

\begin{equation}
\label{tredici}
H=\int_{-L}^L dx \big[Tr (F^2 - A\lambda)\big]
=\sum_{k=-\infty}^{\infty}\big[Tr(a_k a_{-k})- {i\pi \over L}k Tr(a_k
b_{-k})\big].
\end{equation}

When $k\not= 0$ it is natural to introduce
positive and negative frequencies; we define
for $k>0$

\begin{equation}
\label{diciassette}
b_k^{\pm}=b_{\pm k}, \qquad\qquad \pm ia_k^{\pm}=a_{\pm k},
\end{equation}

with the conjugation properties

\begin{equation}
\label{diciotto}
(b_k^{\pm})^{\dag}=b^{\mp}, \qquad\qquad (a_k^{\pm})^{\dag}=a^{\mp}.
\end{equation}

The classical hamiltonian takes the form

\begin{equation}
\label{venti}
H=Tr(a_0^2) + 2\sum_{k=1}^{\infty}Tr(a_k^+ a_k^-)+ {\pi \over L}
\sum_{k=1}^{\infty}k Tr(a_k^+ b_k^- + b_k^+ a_k^-).
\end{equation}

Similarly eq.(\ref{undici}) can be written as

\begin{equation}
\label{ventuno}
\lambda={1\over \sqrt {2L}}\big[\lambda_0+\sum_{k=1}^{\infty}\big(\lambda_k^+
exp(i{\pi k(t+x)\over L})+\lambda_k^-
exp(-i{\pi k(t+x)\over L})\big)\big].
\end{equation}

Thanks to eq.(\ref{sei}), frequencies do not mix under time
evolution. Moreover we get

\begin{equation}
\label{ventidue}
\lambda_0= - {ig \over {2\sqrt{L}}}[b_0,a_0]+{g \over {2\sqrt{L}}}
\sum_{p=1}^{\infty}\big([b_p^-,a_p^+]-[b_p^+,a_p^-]\big),
\end{equation}

\begin{eqnarray}
\label{ventitre}
\lambda_k^-&=&-{\pi k\over L}a_k^- - {ig \over {2\sqrt{L}}}[b_k^-,a_0]
-{g \over {2\sqrt{L}}}[b_0,a_k^-]+
{g \over {2\sqrt{L}}}\sum_{p=1}^{\infty}[b_{p+k}^-,a_p^+]-\nonumber\\
&-&{g \over {2\sqrt{L}}}\sum_{p=1}^{k-1}[b_{k-p}^-,a_p^-]-
{g \over {2\sqrt{L}}}\sum_{p=k+1}^{\infty}[b_{p-k}^+,a_p^-],\\
k&>&0,\qquad\qquad\sum_{p=1}^{0}=0,\nonumber
\end{eqnarray}

and

\begin{equation}
\label{ventiquattro}
\lambda_k^+=(\lambda_k^-)^{\dag}.
\end{equation}

We notice that the hamiltonian describes a free theory; it
does not depend on the coupling constant $g$ and does not
mix Fourier components. Actually the dependence on $g$
is buried in the expression of the Gauss' operator.

The canonical Poisson brackets for our system are \cite{Ba85}

\begin{equation}
\label{quattordici}
\big\{A^r(t,x),F^s(t,y)\big\}= \sqrt2\delta^{rs}\delta(x-y),
\end{equation}

the Dirac distribution $\delta$ being understood as its periodic
generalization

\begin{equation}
\label{quindici}
\delta_P(x-y)={1\over 2L}\sum_{k=-\infty}^{\infty}exp[{i\pi k \over L}
(x-y)].
\end{equation}

Eq.(\ref{quattordici}) entails

\begin{equation}
\label{sedici}
\{b_k^r,a_j^s\}= \delta^{rs}\delta_{k,-j},
\end{equation}

all other brackets vanishing.

For $k\ne 0$ the algebra (\ref{sedici}) becomes

\begin{equation}
\label{diciannove}
\{b_k^{r\mp},a_j^{s\pm}\}=\mp i \delta^{rs}\delta_{kj}.
\end{equation}

At the classical level we can impose the following
residual gauge condition on zero modes

\begin{equation}
\label{gauss}
b_0^{\alpha}\simeq 0,
\end{equation}

for any $\alpha$ not belonging to the Cartan subalgebra.
Here $\simeq$ denotes weakly equal in Dirac's terminology,
as the constraints (\ref{gauss}) are not compatible with the
canonical brackets (\ref{sedici}).

Consistency under time evolution of (\ref{gauss}) entails
\begin{equation}
\label{gauss1}
a_0^{\alpha}\simeq 0,
\end{equation}
for the same values of $\alpha$. Together they give rise to a couple
of ``second class" constraints, that will be enough to get
the vanishing of the first term in the right hand side of eq.
(\ref{ventidue}), in partial fulfilment of the Gauss' law.
To impose the
vanishing of all the zero modes would neither be requested
nor be allowed, as it
will become clear in the following. We should remark that
zero modes do not contribute in the naive
decompactification limit $L \to \infty$. Consequently, in such a
limit, our treatment smoothly tends to the one developed in
\cite{Basgri94}, in spite of the partial use of Gauss' law
we made to cast $b_0$ as in (\ref{gauss}).

Obviously constraints (\ref{gauss}),(\ref{gauss1})
are incompatible with the Poisson
brackets (\ref{sedici}) for $k=0$. Those brackets have to be modified
into Dirac brackets, according to the equation
\begin{equation}
\label{Dirac}
\{b_0^{r},a_0^{s}\}_{\cal D}=\{b_0^{r},a_0^{s}\}-
\{b_0^{r},a_0^{\alpha}\}
\{a_0^{\alpha},b_0^{\beta}\}\{b_0^{\beta},a_0^{s}\},
\end{equation}
where $\alpha$ and $\beta$ do not belong to the Cartan subalgebra.

At a classical level Gauss' law is to be imposed as a constraint
on the allowed trajectories of the system.

Quantization is performed turning Dirac brackets into quantum
commutators. As a consequence zero mode operators not belonging to
the Cartan subalgebra vanish, whereas zero mode operators belonging
to the Cartan subalgebra satisfy canonical commutation relations.

A Fock vacuum $|{\bf \Omega \Big>}$ will then be defined as
the state annihilated by $a_k^-$ and $b_k^-$ for any positive $k$.

We explicitly remark the possibility of having negative norm
states; the excitations we are quantizing have a ghostlike
character \cite{Ba85} and therefore cannot be present in the
Hilbert space of ``physical" states. They will be extruded
by imposing the Gauss' law fulfilment.

The Gauss' law will be imposed at the quantum level in an average
sense by means of the weak condition

\begin{equation}
\label{ventisette}
{\bf \Big<}{\bf\Psi}_{phys}|\lambda^r(t,x)|{\bf \Phi}_{phys}{\bf\Big>}=0,
\end{equation}

which is stable under time evolution, the ``physical" states belonging
to a Hilbert space ${\cal H}_{phys}$
with positive semidefinite metric \cite{Ba85}.
Eq.(\ref{ventisette}) can be Fourier
decomposed, leading to

\begin{equation}
\label{ventotto}
{\bf \Big<}{\bf\Psi}_{phys}|\lambda_0|{\bf \Phi}_{phys}{\bf \Big>}=0,
\end{equation}

and

\begin{equation}
\label{ventinove}
\lambda_k^-|{\bf \Phi}_{phys}{\bf \Big>}=0,\qquad\qquad \forall k>0.
\end{equation}

Eqs.(\ref{ventotto}) and (\ref{ventinove})
can in turn be realized by requiring

\begin{equation}
\label{trenta}
a_k^-|{\bf \Phi}_{phys}{\bf \Big>}=0,\qquad\qquad
b_k^-|{\bf \Phi}_{phys}{\bf \Big>}=0,\qquad\qquad \forall k>0.
\end{equation}

The hamiltonian, when restricted to the ``physical" subspace,
just becomes

\begin{equation}
\label{trentuno}
H_{phys}=Tr(a_0^2)
\end{equation}

and we are left with only zero modes belonging to the Cartan subalgebra.

In the following we shall focus our interest on $su(2)$ and on $su(3)$.

In $su(2)$ we choose $b_0$ along $\sigma^3$: $b_0=\beta_3 {\sigma^3
\over 2}$.
At a classical level, we have the possibility of performing
the residual gauge
transformation

\begin{eqnarray}
\label{trentatre}
\big[U(t+x)\big]&=&exp\big[{{i\pi } \over L}(t+x)\,\sigma^3 \big]
\nonumber\\
&=&cos\big[{\pi  \over L}(t+x)\big]+\,i\,\sigma^3
\,sin\big[{\pi \over L}(t+x)\big],
\end{eqnarray}
which is globally defined on the cylinder.
It induces on $\beta_3$ the transformation

\begin{equation}
\label{trentaquattro}
\beta_3\to \beta_3+{{4\pi}\over {g\sqrt L}}
\end{equation}

suggesting the introduction of an angular variable $\theta$
according to the equation

\begin{equation}
\label{trentaquattrobis}
\beta_3=2\,\theta({g\sqrt L})^{-1},\qquad\qquad -\pi \leq \theta \leq \pi.
\end{equation}

The operator U, which generates the translation of $\beta_3$ of a period,
has the following quantum representation

\begin{equation}
\label{quantum}
U= exp [{{4 \pi i}\over {g\sqrt L}} \alpha_3],
\end{equation}
where $a_0=\alpha_3 {\sigma^3\over 2}$.
The canonical algebra suggests

\begin{equation}
\label{trentaquattroter}
\alpha_3=-\,i\,{{g\sqrt L}\over 2}{d\over {d\theta}}.
\end{equation}

The corresponding spectrum of the hamiltonian in
eq.(\ref{trentuno}), when restricted to the ``physical" subspace,
turns out to be
\cite{He88}

\begin{equation}
\label{trentanovebis}
E_n={g^2\,L\over 8}\,n^2,
\end{equation}

(see eq.(\ref{trentaquattroter})).

We remark that, apart from the fundamental level $n=0$,
all other energy eigenvalues are linearly increasing with
$L$ and would diverge in the decompactification
limit. They can be generated by acting on the fundamental
level by the operator $exp[\,i\,\theta]$, which in turn
is nothing but the average on the state $|{\bf \Omega \Big>}$
of the Wilson loop wrapping once around the cylinder at
$t=0$.

A unique vacuum with respect to the Hamiltonian
corresponds to $n=0$
\cite{He88}. The situation might be quite different in the presence
of dynamical fermions.

The space of states can be taken as the direct
product of the Fock space related to non vanishing frequencies times
the Hilbert space of zero modes ${\cal R}$.

The ``physical" Hilbert space ${\cal H}_{phys}$ is realized as
the product $|{\bf \Omega \Big>} \times {\cal R}$. Nevertheless
we shall discover that the ghost-like degrees of freedom, although
excluded from ${\cal H}_{phys}$, entail important consequences even
on gauge invariant quantities (typically Wilson loops).

It is now instructive to see how this treatment is generalized
to su(3).

Let us expand $b_0$ on the Cartan subalgebra of su(3) that we
parametrize according to the Gell-Mann basis

\begin{equation}
\label{trentaquattroIV}
b_0=\beta_3T^3+\beta_8T^8.
\end{equation}

Then, following eq.(\ref{gauss1}) it is natural to decompose

\begin{equation}
\label{trentacinque}
a_0=\alpha_3T^3+\alpha_8T^8.
\end{equation}

We have still the possibility of performing residual gauge
transformations of the kind

\begin{equation}
\label{trentaseiter}
U(t+x)=exp\big[{{2\,i\pi } \over L}(t+x)\,n_3T^3\big]\,\,
exp\big[{{i\pi } \over L}(t+x)\,n_8(T^3+\sqrt 3 \, T^8)\big],
\end{equation}

which are globally defined on the cylinder as long as $n_3$ and $n_8$
take integral values.

As a consequence

\begin{eqnarray}
\label{trentasette}
\theta_3&=&{g\sqrt L \over 2}(\beta_3-{1\over{\sqrt 3}}\beta_8),\nonumber\\
\theta_8&=&{g\sqrt L\over{\sqrt 3}}\beta_8
\end{eqnarray}

turn out to be angular variables taking values between $-\pi$ and $\pi $.

The algebra in eq.(\ref{Dirac}) suggests the expressions

\begin{eqnarray}
\label{trentottobis}
\alpha_3&=&{g\sqrt L \over 2}{\cal L}_3,\nonumber\\
\alpha_8&=&-{g\sqrt L\over{2\sqrt 3}}[{\cal L}_3- 2{\cal L}_8],
\end{eqnarray}

${\cal L}_3$ and ${\cal L}_8$ being the angular momenta
conjugate to $\theta_3$ and $\theta_8$ respectively.

Eqs.(\ref{trentuno}) and (\ref{trentottobis}) then provides
the spectrum of the hamiltonian
\begin{equation}
\label{trentanove}
E_{n_3,n_8}= {g^2 L\over 6} (n_3^2-n_3n_8+n_8^2),\qquad\qquad n_3,n_8\in
{\cal Z}.
\end{equation}

We remark that, apart from the fundamental level
$(n_3=n_8=0)$, all other energy eigenvalues are linearly increasing with
$L$ and would diverge in the decompactification
limit. A unique vacuum corresponds to $n_3=n_8=0$
\cite{He88}.

\section{Wilson loops}
\label{terza}
\vskip 0.3truecm
Recently perturbative
Wilson loop calculations in 1+1 dimensions gave different results
according to different expressions for the propagator \cite{Basgri94}.
As those results have been obtained in the continuum, we feel
important to reexamine them on the cylinder, also in view of
the recent abundant literature on $QCD_2$.

In order to avoid an immediate interplay with topological features, we
consider a Wilson loop entirely contained in the basic interval
$-L\le x \le L$. We choose a rectangular
Wilson loop $\gamma$ with light--like sides, directed along the vectors
$n_\mu$ and $n^*_\mu$, with lengths $\lambda$ and $\tau$
respectively, and parametrized according to the equations:

\begin{eqnarray}
\label{quarantasei}
C_1:x^\mu (s) &=& n^{\mu} {\lambda} s, \nonumber\\
C_2:x^\mu (s) &=& n^{\mu} {\lambda}+ n^{* \mu}{\tau} s, \nonumber\\
C_3:x^\mu (s) &=& n^{* \mu}{\tau} + n^{\mu} {\lambda}( 1-s), \\
C_4:x^\mu (s) &=& n^{* \mu}{\tau} (1 - s), \qquad 0 \leq s \leq 1, \nonumber
\end{eqnarray}
with ${\lambda + \tau}<2\sqrt 2 L$.

We are interested in the quantity

\begin{equation}
\label{quarantasette}
W(\gamma)= {\bf \Big< 0}|Tr{\cal T}{\cal P}\Big( exp\Big[ig\oint_{\gamma}
A dx^+\Big]\Big)|{\bf 0 \Big>},
\end{equation}

where ${\cal T}$ means time-ordering and
${\cal P}$ color path-ordering along $\gamma$.

The vacuum state belongs to the physical Hilbert space as far as
the non vanishing frequency parts are concerned; it is indeed the
Fock vacuum $|{\bf \Omega \Big>}$. Then we
consider its direct product with the lowest eigenstate of the
Hamiltonian in eq.(\ref{trentuno}) concerning zero modes. This
eigenstate belongs to the domain of $a_0$; however, when $b_0$
acts on it, it generates new states which are no longer in
the domain of $a_0$, as is well known from elementary quantum
mechanics.
As a consequence, due to zero modes, we cannot define a ``bona fide"
propagator for our theory: this should not come to a surprise
in view of the topological nature of $b_0$.
On the other hand a propagator is not required in
eq.(\ref{quarantasette}).

We shall first discuss
the simpler case of QED, where no color ordering is involved.
Eq.(\ref{quarantasette}) then becomes

\begin{equation}
\label{quarantotto}
W(\gamma)= {\bf \Big< 0}|{\cal T}\Big( exp\Big[ig\oint_{\gamma}
A dx^+\Big]\Big)|{\bf 0 \Big>},
\end{equation}

and a little thought is enough to realize the factorization property
\begin{equation}
\label{quarantanove}
W(\gamma)= {\bf \Big< 0}|{\cal T}\Big( exp\Big[{ig \over {\sqrt
{2L}}}\oint_{\gamma}
(b_0 + a_0t) dx^+\Big]\Big)|{\bf 0 \Big>}{\bf \Big< 0}|{\cal T}
\Big( exp\Big[ig \oint_{\gamma}
\hat{A}(t,x) dx^+\Big]\Big)|{\bf 0 \Big>} = W_0\cdot\hat{W},
\end{equation}
according to the splitting in eq.(\ref{dieci}).
In turn the Wilson loop $\hat{W}$ can also be expressed as a Feynman
integral starting
from the lagrangian in eq.(\ref{uno}) for QED, without the zero mode

\begin{equation}
\label{cinquanta}
\hat{W}(\gamma)={\cal N}^{-1} \Big( exp\Big[g\oint_{\gamma}{\partial
\over {\partial J}}dx^+\Big]\Big)
\Big[\int {\cal D} \hat{A}\, {\cal D}\lambda\, exp\,\,i\Big(\int d^2x({\cal L}+
J\hat{A})\Big)\Big]_{\Big| J=0},
\end{equation}

${\cal N}$ being a suitable normalization factor.

Standard functional integration gives

\begin{eqnarray}
\label{cinquantuno}
\hat{W}(\gamma)&=&{\cal N}^{-1}\Big(exp\Big[g\oint_{\gamma}{\partial
\over {\partial J}}dx^+\Big]\Big) \nonumber\\
&& exp\Big[i\,\int\!\!\!\int d^2\xi d^2\eta
J(\xi)\hat{G}_c(\xi-\eta)
J(\eta)\Big]_{\Big| J=0}.
\end{eqnarray}

Using the equations of motion, one can
verify that $\hat{G}_c$
obeys the free inhomogeneous hyperbolic differential equation

\begin{equation}
\label{cinquantadue}
\partial^2_- \hat{G}_c(t,x)= \delta(t) [\delta(x) - {1\over{2L}}],
\end{equation}

with causal boundary conditions.

The canonical algebra provides us with the solution

\begin{equation}
\label{cinquantatre}
\hat{G}_c(t,x)={|t|\over {2L}}\Big[\theta(t)\sum_{n=1}^{\infty}
exp\big({-i\pi n\over L}(t+x)\big)+\theta(-t)\sum_{n=1}^{\infty}
exp\big({i\pi n\over L}(t+x)\big)\Big].
\end{equation}

The causal nature of positive and negative frequency treatment is
clearly exhibited.

Eq.(\ref{cinquantatre}) should be compared to the expression
for the ``potential" used by 't Hooft \cite{Ho74}

\begin{equation}
\label{cinquantacinque}
\hat{G}(t,x)={|t|\over {4L}}\sum_{n=-\infty,n\ne 0}^{\infty}
exp\big({i\pi n\over L}(t+x)\big)={1\over 2} |t| [\delta_P(t+x)-
{1\over{2L}}],
\end{equation}

which coincides with its real part. Eq.(\ref{cinquantacinque}) is
indeed the Green's function one obtains when interpreting
$\partial_-^2$ not as a propagation, but rather as a constraint.
One can also easily check
the relation

\begin{equation}
\label{cinquantasei}
\hat{G}_c(t,x)=\hat{G}(t,x)-{i\,t\over{4L}}P\,ctg[{\pi\over{2L}}(t+x)],
\end{equation}

where $P$ means Cauchy principal value.
Eq.(\ref{cinquantasei}) is the explicit
causal expression of the integral operator $\hat{G}_c$. It
looks like a complex ``potential" kernel, the absorptive part being
related to the presence of ghost-like excitations, which are
essential to recover the M-L prescription in the decompactification
limit $L \to \infty$

\begin{equation}
\label{cinquantaquattro}
\hat{G}_c \to -{i\,t\over{2\pi}}{1\over
{(t+x)-i\epsilon\, sign(t)}}= -{i\,t\over{2\pi}}{1\over
{(t+x)-i\epsilon\, sign(t-x)}}.
\end{equation}

We remark that zero
modes are irrelevant in the limit $L \to \infty$.

Now, introducing eq.(\ref{cinquantasei}) in eq.(\ref{cinquantuno}),
we get the expression

\begin{eqnarray}
\label{cinquantasette}
\hat{W}(\gamma)&=&exp\Big[ i\,g^2 \oint_{\gamma}dx^+\,\oint_{\gamma}
dy^+\,\hat{G}_c(x^+-y^+,x^-\,-y^-)\Big]\nonumber\\
&=&exp\Big[ i\,g^2 \oint_{\gamma}dx^+\,\oint_{\gamma}
dy^+\,\hat{G}(x^+-y^+,x^-\,-y^-)\Big]\nonumber\\
&=& exp\Big[-i\,{g^2\,{\cal A} \over 2}\Big]\,
exp\Big[ -i\,g^2 \oint_{\gamma}dx^+\,\oint_{\gamma}
dy^+\,{|x^+ + x^- -y^+ - y^-|\over {4L \sqrt 2}}\Big]
\end{eqnarray}

the absorptive part of the ``potential" averaging to zero. The
quantity ${\cal A}=\lambda \tau$ is the area of the loop.
The same result can also be obtained by operatorial
techniques, using Wick's theorem and the canonical algebra.

We are thereby left with the problem of computing $W_0$

\begin{equation}
\label{cinquantotto}
W_0(\gamma)= {\bf \Big< 0}|{\cal T}\Big( exp\Big[{ig \over {\sqrt
{2L}}}\oint_{\gamma}
(b_0 + a_0t) dx^+\Big]\Big)|{\bf 0 \Big>}.
\end{equation}

The expansion of the exponential in (\ref{cinquantotto}) is delicate
and, if not carefully treated, could give rise to ambiguous results.
We recall that, when $b_0$ acts on the vacuum state, it generates a
state no longer belonging to the domain of $a_0$. This problem can
be overcome in the following way.
Using the canonical algebra, the ${\cal T}$-product of $n$ factors
in the expansion of (\ref{cinquantotto}),
can be recursively ordered, by generalizing Wick's theorem,
in a sum of operators in which $a_0$ factors appear on the right
or do not appear at all. Terms without $a_0$ are in turn of two kinds:
either they exhibit some $b_0$'s or are $c$-number products of
contractions. Only those last terms provide non vanishing
contributions. As a matter of fact terms with unpaired $b_0$'s
vanish upon the related $\oint_{\gamma} dx^+$-integration already
at the operator level, whereas terms
containing $a_0$'s vanish when acting on the vacuum on the right
\cite{Man85}.

Contraction terms are unambiguous and obviously entail an even
number of fields; thus the $2n$-th term of the expansion
provide us with $(2n-1)!!$ expressions of the kind

\begin{equation}
\label{cinquantanove}
{ig^2 \over {2L}}\oint_{\gamma}dx_{j}^+\oint_{\gamma}
dx_{k}^+\,max\{t_{j},t_{k}\}= i\,g^2 \oint_{\gamma}dx^+\,\oint_{\gamma}
dy^+\,{|x^+ + x^- -y^+ - y^-|\over {4L \sqrt 2}}.
\end{equation}

Summing over $n$ we recover exponentiation with a factor which
{\it exactly cancels} the last exponential in eq.(\ref{cinquantasette}),
leaving the pure loop area result, only as a consequence of the
canonical algebra, and even in the presence of
a topological degree of freedom. The result exactly coincides
with the one we would have obtained introducing in eq.
(\ref{cinquantasette}) the complete Green's function, $i.e.$ with
the zero mode included.

However eq.(\ref{cinquantasette}) unfortunately is unable to
discriminate between the two different Green's functions $\hat {G}_c$
and $\hat {G}$.

The generalization to the non abelian case is far from being trivial,
owing to an intertwining between space and group algebra variables.
In particular topological excitations get inextricably mixed with
the frequency parts. The very idea of combining zero mode contribution
and frequency contractions into an ``effective" propagator is
unjustified in our opinion. Yet it is true that in the limit $L \to
\infty$ zero modes do not contribute.
Should we close our eyes and try to evaluate
the Wilson loop by summing perturbative exchanges in the form
$G(t,x)={1\over 2} |t| \delta_P(t+x)$, as suggested by the QED case,
eq.(\ref{quarantasette}) could be explicitly
evaluated, thanks to the ``delta-like" character of the ``potential".
We would indeed obtain

\begin{eqnarray}
\label{eresia1}
W(\gamma)&=&Tr\,{\cal P}\Big(exp\Big[g\int_{C_4}{\partial
\over {\partial J^+}}dx^+\Big]\Big)
\,{\cal P}\Big(exp\Big[g\int_{C_2}{\partial
\over {\partial J^+}}dx^+\Big]\Big) \nonumber\\
&& exp\Big[i\,Tr\int\!\!\!\int d^2\xi d^2\eta
J^+(\xi)G(\xi-\eta)
J^+(\eta)\Big]_{\Big| J^+=0},
\end{eqnarray}

namely

\begin{equation}
\label{eresia2}
W(\gamma)=Tr\,{\cal P}\Big(exp\Big[g\int_{C_4}{\partial
\over {\partial J}}dx\Big]\Big)
\,{\cal P}\Big(exp\Big[{i\,g\,\lambda \over 2} \int_{C_4}
J\,dx\Big]\Big)_{\Big| J=0},
\end{equation}

and finally

\begin{equation}
\label{eresia3}
W(\gamma)= exp\Big[-i\,{g^2\,N\,C_F\,\lambda\,\tau \over 2}\Big],
\end{equation}

$C_{F}$ being the quadratic Casimir of the fundamental representation of
$su(N)$. The area (${\cal A}=\lambda\,\tau$) law behaviour of
the Wilson loop we have found in this case together with the occurence
of a simple exponentiation in terms
of the Casimir of the fundamental representation, is a quite
peculiar result, insensitive to the decompactification limit
$L\to \infty$. It is rooted in the particularly simple expression
for the ``potential" we have used that coincides with the one
often considered in analogous Euclidean calculations \cite{bra80}.

However canonical quantization suggests that we should rather
use the propagator $G_{c}(t,x)={1\over 2} |t| \delta_P(t+x)-
{i\,t\over{4L}}P\,ctg[{\pi\over{2L}}(t+x)]$ (see
eq.(\ref{cinquantasei})).
Then a full resummation of perturbative exchanges would be
no longer possible,
owing to the presence of non vanishing cross diagrams.
Already at ${\cal O}(g^4)$, a tedious but straightforward calculation
of the sum of all the ``cross" diagrams  in fig. 1 leads to the result

\begin{equation}
\label{follia1}
W_{cr}=({g^2 \over {4\pi}})^2\, 2\,C_F\,C_A ({\cal A})^2 \int_{0}^{1}
d\,\xi \int_{0}^{1} d\,\eta\,log{|sin\rho(\xi-\eta)|\over
{|sin\rho\xi|}}log{|sin\rho(\xi-\eta)|\over
{|sin\rho\eta|}},
\end{equation}

where $\rho= {\pi\lambda\over{\sqrt 2 L}}$.
One immediately recognizes the appearance of the
quadratic Casimir of the adjoint
representation ($C_A$); moreover, besides the area dependence, a
dimensionless parameter $\rho$, which measures the ratio of the loop
length $\lambda$ to the interval length $L$, explicitly occurs.
In the decompactification limit $\rho\to 0$, the expression \cite{Basgri94}

\begin{equation}
\label{follia2}
W_{cr}=({g^2 \over {4\pi}})^2\, C_F\,C_A ({\cal A})^2 {\pi^2\over 3}
\end{equation}

is smoothly recovered. While the presence of zero modes puts severe
doubts on any perturbative calculation owing to the poor definition
of propagators, it is perhaps not surprising that in the
limit $L \to \infty$
the perturbative result in the continuum is correctly reproduced.

\section{Concluding remarks}
\label{quarta}
\vskip 0.3truecm
To conclude, we briefly summarize our results.
We have succeeded in canonically quantizing at equal ${\bf time}$
in light-cone gauge pure Yang-Mills theories, defined on the
space-time $S_1\times R$ cylinder. Gauss' law is imposed in a weak sense,
e.g. as a condition on the Fock space vectors. Creation and
annihilation operators, corresponding to non vanishing frequencies,
do not contribute in the ``physical" Hilbert space; nevertheless
they give rise to a propagator, which is the light-cone counterpart
of the Coulomb interaction term in the Coulomb gauge. Equal time
quantization induces a causal behaviour of this propagator, making it
different from the ``contact" expression often considered in the
literature. In the decompactification limit $L \to \infty$,
it naturally becomes the causal M-L distribution.

Zero modes are present as topological degrees of freedom and
quantized according to their winding around the cylinder.

A Wilson loop with light-like sides is exactly computed in QED;
the area law is recovered even in the presence of topological
degrees of freedom.

In the non Abelian case a perturbative approach would be
hindered by the poor definition of a propagator in the presence
of topological excitations. In addition the causal nature of the
sum over frequency contributions prevents an exact
solution; a perturbative solution at ${\cal {O}}(g^4)$ in the
limit $L \to \infty$
reproduces smoothly the result we have previously found in the
continuum \cite{Basgri94}.

Another way of understanding the difficulties of the Wilson loop
calculation is to observe its equivalence with a fermionic problem:
it is well-known \cite{Gervi79} that the two-point function for dynamical
one-dimensional fermions living on a loop and interacting with a Yang-Mills
theory, defined on a manifold containing the loop itself, {\it exactly} gives
(at coincident points) the value of the related Wilson loop
for pure Y-M theory. In this sense, due to the presence of interactions
between these auxiliary fermions, frequencies come into game, mixing with
the (topological) gauge invariant degrees of freedom. We remark that the
treatment of loops winding around the cylinder is somewhat easier to handle
than of the non-wrapped ones because they
are directly related to the quantum mechanical degrees of fredoom: large-N
computations for such a class of observables are presented in \cite{Gross94}.
On the other hands it was shown in \cite{Poly94} that the presence of
Wilson lines (described by the formalism of one-dimensional fermions)
turns the theory into an effective Calogero-Sutherland model \cite{Calo69}:
it could be very insightful to understand if in our case the exact evaluation
of zero-modes contributions to the Wilson loop corresponds to the solution
of some suitable quantum mechanical model.

Other interesting issues are open to future investigation, the
most exciting one perhaps concerning the introduction of dynamical
(two-dimensional) massless fermions.

Finally it should be worth understanding to what extent the
features we have found on the 1+1 cylinder, may apply to the more
difficult , but realistic case of 3+1 dimensions.
\vfill\eject

\newpage

\begin{picture}(450,90)
\put(0,0){
\begin{picture}(70,90)
\put(0,0){\makebox(0,0){(0,0)}}
\put(60,0){\makebox(0,0){($\tau$,0)}}
\put(60,80){\makebox(0,0){($\tau$,$\lambda$)}}
\put(0,80){\makebox(0,0){(0,$\lambda$)}}
\thicklines
\put(10,10){\framebox(40,60)}
\put(60,50){\vector(0,-1){20}}
\thinlines
\put(20,10){\line(1,3){20}}
\put(20,70){\line(1,-3){20}}
\end{picture}}

\put(150,0){
\begin{picture}(70,90)
\put(0,0){\makebox(0,0){(0,0)}}
\put(60,0){\makebox(0,0){($\tau$,0)}}
\put(60,80){\makebox(0,0){($\tau$,$\lambda$)}}
\put(0,80){\makebox(0,0){(0,$\lambda$)}}
\thicklines
\put(10,10){\framebox(40,60)}
\put(60,50){\vector(0,-1){20}}
\thinlines
\put(30,70){\oval(20,10)[t]}
\put(30,70){\line(0,-1){60}}
\end{picture}}

\put(300,0){
\begin{picture}(70,90)
\put(0,0){\makebox(0,0){(0,0)}}
\put(60,0){\makebox(0,0){($\tau$,0)}}
\put(60,80){\makebox(0,0){($\tau$,$\lambda$)}}
\put(0,80){\makebox(0,0){(0,$\lambda$)}}
\thicklines
\put(10,10){\framebox(40,60)}
\put(60,50){\vector(0,-1){20}}
\thinlines
\put(30,10){\oval(20,10)[b]}
\put(30,70){\line(0,-1){60}}
\end{picture}}
\end{picture}

\bigskip
\bigskip
\bigskip

\begin{picture}(450,90)
\put(80,0){
\begin{picture}(70,90)
\put(0,0){\makebox(0,0){(0,0)}}
\put(60,0){\makebox(0,0){($\tau$,0)}}
\put(60,80){\makebox(0,0){($\tau$,$\lambda$)}}
\put(0,80){\makebox(0,0){(0,$\lambda$)}}
\thicklines
\put(10,10){\framebox(40,60)}
\put(60,50){\vector(0,-1){20}}
\thinlines
\put(25,10){\oval(20,10)[b]}
\put(35,10){\oval(20,10)[t]}
\end{picture}}

\put(230,0){
\begin{picture}(70,90)
\put(0,0){\makebox(0,0){(0,0)}}
\put(60,0){\makebox(0,0){($\tau$,0)}}
\put(60,80){\makebox(0,0){($\tau$,$\lambda$)}}
\put(0,80){\makebox(0,0){(0,$\lambda$)}}
\thicklines
\put(10,10){\framebox(40,60)}
\put(60,50){\vector(0,-1){20}}
\thinlines
\put(25,70){\oval(20,10)[t]}
\put(35,70){\oval(20,10)[b]}
\end{picture}}
\end{picture}

\bigskip
\bigskip
\bigskip

\begin{picture}(450,20)
\put(170,0){{\bf Fig. 1}}
\end{picture}

\bigskip
\bigskip
\bigskip
\bigskip
\bigskip
{FIGURE CAPTIONS}
\bigskip

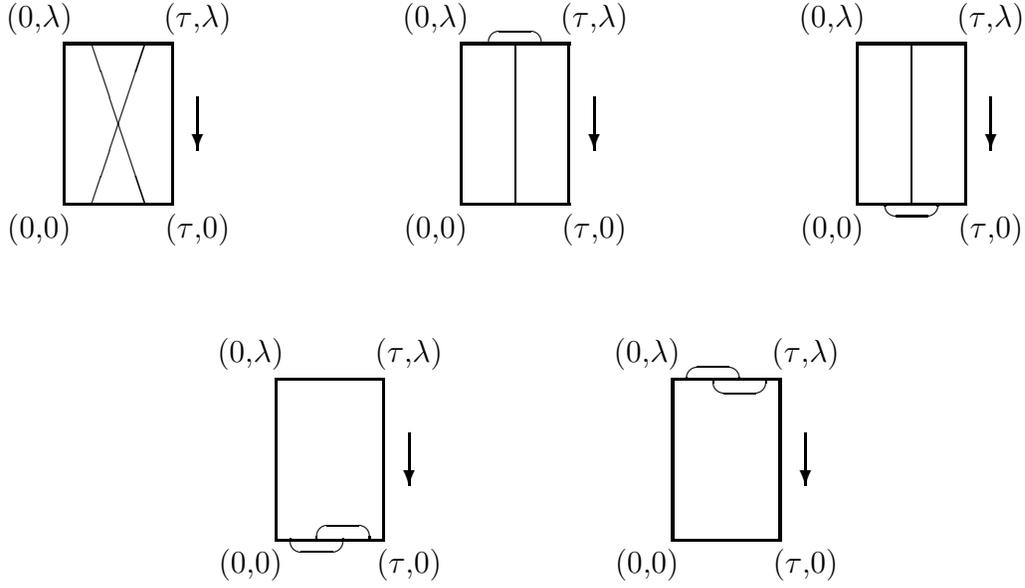
{\bf Fig. 1:} Crossed diagrams contributing with a $C_FC_A$ term to the $g^4$
perturbative evaluation of the Wilson loop with causal propagator $G_c(t,x)$.
Thick lines denote the rectangular loop in the $x^+ \times  x^-$ plane, with
sides length $\tau$ and $\lambda$, respectively. Thin lines denote gluon
propagators.


\begin{thebibliography}{100}
\bibitem{Man85}{Obviously this right ordering can be replaced by a
left ordering without affecting the result for $W_0(\gamma)$.}
\end{thebibliography}
\end{document}